\documentclass[preprint,amssymb,amsmath]{revtex4}
\usepackage[dvipdfm,colorlinks,hypertex]{hyperref}
\usepackage{graphicx}  
\usepackage{bm}        

\begin{document}

\title{Site-specific symmetry sensitivity of angle-resolved photoemission spectroscopy in layered palladium diselenide
}
\author{M.~Cattelan}
\affiliation{Elettra Sincrotrone Trieste, SS 14, Km 163.5 AREA Sci. Pk., 34149 Trieste, Italy} 
\author{C.~J.~Sayers}
\affiliation{Dipartimento di Fisica, Politecnico di Milano, 20133 Milano, Italy}
\author{D.~Wolverson}
\affiliation{University of Bath, Ctr. Nanosci. and Nanotechnol., Dept. Phys., Bath BA2 7AY, UK}

\author{E.~Carpene}
\email{ettore.carpene@polimi.it}
\affiliation{IFN-CNR, Dipartimento di Fisica, Politecnico di Milano, 20133 Milano, Italy}

\begin{abstract}

Two-dimensional (2D) materials with puckered layer morphology are promising candidates for next-generation opto-electronics devices owing to their anisotropic response to external perturbations and wide band gap tunability with the number of layers. Among them, PdSe$_2$ is an emerging 2D transition-metal dichalcogenide with band gap ranging from $\sim 1.3$~eV in the monolayer to a predicted semimetallic behavior in the bulk. Here we use angle-resolved photoemission spectroscopy to explore the electronic band structure of PdSe$_2$ with energy and momentum resolution. Our measurements reveal the semiconducting nature of the bulk. Furthermore, constant binding-energy maps of reciprocal space display a remarkable site-specific sensitivity to the atomic arrangement and its symmetry. Supported by density functional theory calculations, we ascribe this effect to the inherent orbital character of the electronic band structure. These results not only provide a deeper understanding of the electronic configuration of PdSe$_2$, but also establish additional capabilities of photoemission spectroscopy.

\end{abstract}

\maketitle

Transition metal dichalcogenides (TMDs) host highly attractive properties for fundamental studies of novel physical phenomena and for applications ranging from opto-electronics to sensing at the nanoscale \cite{tmd1, tmd2}. Among all TMDs, those based on noble metals (Pd, Pt) have received less attention because of their high cost, until the recent discovery of a layer-controllable metal-to-semiconductor transition \cite{oye, noble1,noble2,noble3}, which motivated their investigation in the last few years. Similar to the extensively investigated black phosphorus \cite{bp1,bp2,bp3,bp4}, with a band gap varying from 0.3~eV in the bulk to 1.5~eV in the monolayer \cite{bpgap}, PdSe$_2$ is characterized by an in-plane puckered structure resulting in an anisotropic response to external stimuli, such as light \cite{shg}, electric field \cite{field} and strain \cite{strain,strain2}. In addition, a linear dependence of the band gap with the number of layers has been observed \cite{oye,gap2}: the monolayer is predicted to have an indirect gap of 1.3~eV \cite{omega} which monotonically decreases to 0~eV as the thickness exceeds $40-50$ layers, suggesting semimetallic behavior. However, unlike black phosphorous and the majority of TMDs which share hexagonal crystal structures, the low-symmetry pentagonal atomic arrangement of PdSe$_2$ gives rise to exotic thermoelectric, mechanical and optical properties \cite{penta1,penta2,penta3}.

Here, we investigate the electronic structure of PdSe$_2$ for the first time by angle-resolved photoemission spectroscopy (ARPES). We clarify the semiconducting nature of the bulk by direct measurements of the electronic bands in reciprocal space. Furthermore, we reveal a previously unexplored sensitivity of photoemission to the site-specific crystal symmetry. In particular, constant binding-energy cuts of the surface-projected Brillouin zone (BZ) disclose the dominant chemical/orbital character of the metal atoms on the top-most valence band, while the effect of the chalcogen species becomes relevant at binding energies exceeding 1~eV. This finding is corroborated by plane wave density functional theory (DFT) calculations employing Perdew-Burke-Ernzerhof (PBE) functional.

\begin{figure}[htb]
\includegraphics[width=130mm]{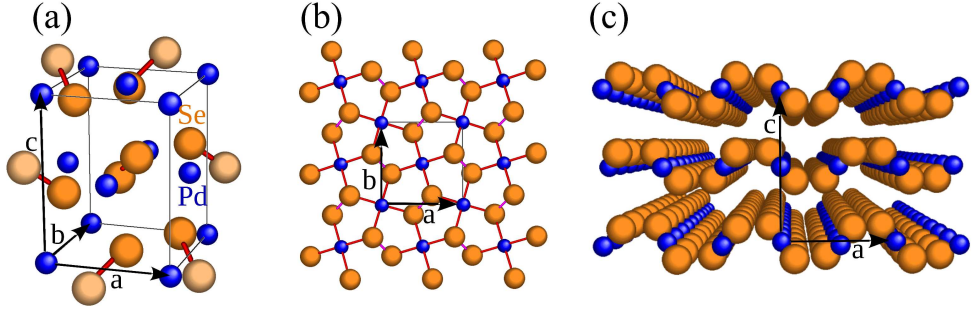}
\caption{Unit cell (a), top (b) and side (c) views of PdSe$_2$ crystal structure. The red links in panel (a) indicate the bonds related to Se$_2$ dimerization.}
\label{f1}
\end{figure}
The crystallographic morphology of PdSe$_2$ is sketched in Fig.\ref{f1}. Its stable configuration is orthorhombic with Pbca space group (\#61) and experimental lattice parameters $a = 0.575$~nm, $b = 0.587$~nm and $c = 0.77$~nm \cite{pbca}, see Fig.\ref{f1}a. Layers are normal to the c-axis. The top view (panel b) shows the characteristic pentagonal atomic arrangement of the monolayer, while the side view (panel c) reveals its puckered structure. Each PdSe$_2$ layer is formed by three atomic planes: Pd atoms in the middle are covalently bound to four Se atoms located on the top and bottom sub-layers. In contrast with other TMDs where the metal atom has $+4$ oxidation state, PdSe$_2$ adopts the $+2$ oxidation \cite{kem}. This is achieved via (Se$_2$)$^{2-}$ dimerization (see red links in Fig.\ref{f1}a) that provides further bonding between the top and the bottom chalcogen sub-layers. 

\begin{figure}[htb]
\includegraphics[width=130mm]{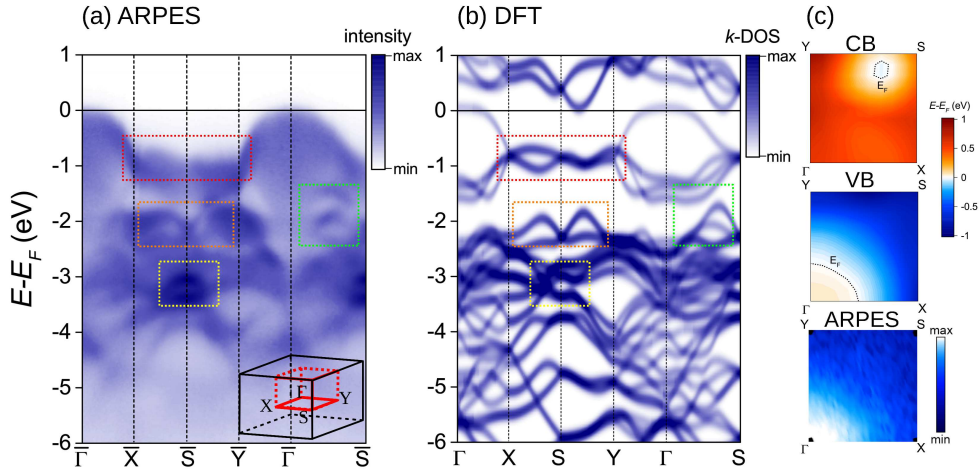}
\caption{(a) Electronic band structure of PdSe$_2$ measured by ARPES ($h\nu = 74$~eV). The inset shows a sketch of the first BZ. (b) $k$-resolved DOS obtained with DFT calculations along the same path in reciprocal space used in panel a. (c) DFT mapping of the CB (top) and VB (middle) on a 2D portion of the first BZ: dashed black lines mark the Fermi level $E_F$. The 2D ARPES intensity map at $E_F$ (bottom) reveals no signature of the CB.}
\label{f2}
\end{figure}

\begin{figure}[htb]
\includegraphics[width=80mm]{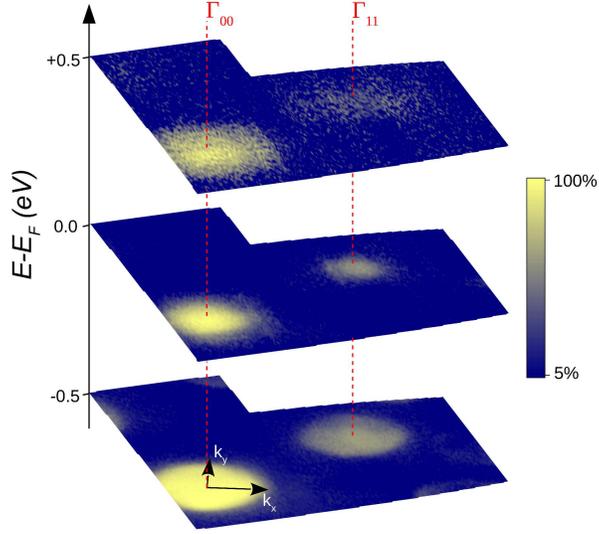}
\caption{ARPES constant energy maps at $E-E_F=-0.5$ (bottom), 0 (middle) and $+0.5$ eV (top), revealing no sign of the CB.}
\label{s3}
\end{figure}

Fig.\ref{f2}a shows the measured band structure of PdSe$_2$ along selected high symmetry directions, following the orthorhombic unit cell notation shown in the inset (see Methods for experimental details). The top of the valence band (VB) is located at the center of the BZ and is adjacent to the Fermi level $E_F$. The large spectral weight at the $S$ point ($E-E_F\simeq-3$~eV) suggests the intersection of several bands.
Fig.\ref{f2}b reports the $k$-resolved density of states ($k$-DOS) obtained with DFT calculations along the same symmetry lines adopted in panel a (see Methods for computational details). The main features of the occupied states are reproduced with remarkable adherence to the experimental data, in particular the dispersion of the top VB and the large DOS at the $S$ point and $E-E_F\simeq-3$~eV, that in fact hosts an intricate overlap of several electronic bands. The coloured rectangles in panels a and b highlight additional features common to both measured and calculated band structures. Our DFT analysis identifies PdSe$_2$ as a semimetal: the top graph in Fig.\ref{f2}c shows the false-colour energy landscape of the lowest conduction band (CB) on a 2D portion of the BZ. The minimum is located close to the $YS$ line (internal coordinates $[k_x,k_y]\simeq[0.33, 0.43]$ at energy $E-E_F\simeq-25$~meV). The middle graph shows the energy landscape of the highest VB, where its maximum occurs at the $\Gamma$ point and energy $E-E_F\simeq+55$~meV. Hence, there is an overall negative gap of approximately 80~meV. In contrast to the calculations, no evidence of electron pockets due to the CB crossing the Fermi level has been observed experimentally, as shown in the lowest graph of Fig.\ref{f2}c that reports the surface-projected ARPES map at the Fermi level. Thus, we conclude that PdSe$_2$ is most likely a semiconductor in the bulk (See Methods - Fig.\ref{s3} - for additional evidence). This result is in agreement with recent optical measurements \cite{opt1,opt2}, but diverges from calculations that predict semimetallic behaviour \cite{oye,omega,nscale,hami}. While it is known that DFT generally underestimates band gaps in semiconductors \cite{dftgap,dftgap2}, the fact that VB and CB are well separated in reciprocal space preserves their individual electronic features, regardless of the computed value of the band gap. In particular, the occupied states (experimentally observed by ARPES) are reproduced by DFT with remarkable accuracy.

\begin{figure}[htb]
\includegraphics[width=130mm]{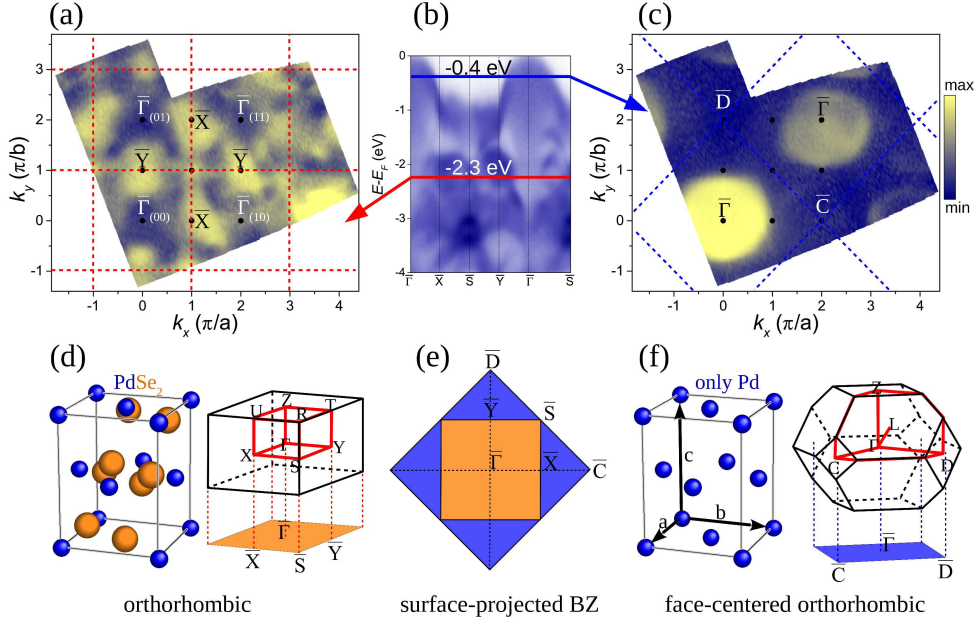}
\caption{Constant binding-energy ARPES maps at (a) $E-E_F = -2.3$~eV and (c) $E-E_F = -0.4$~eV. Red and blue dashed lines correspond to the surface-projected orthorhombic and face-centered orthorhombic (fco) BZs, respectively (see panels d-f). (b) The selected energies are shown on the measured band structure. (d) Orthorhombic unit cell of PdSe$_2$ with the corresponding BZ. The orange-shaded area sketches the surface-projected BZ (SBZ). (f) fco unit cell describing the Pd arrangement with the corresponding BZ and its surface projection (blue-shaded area). (e) Comparison of orthorhombic and fco SBZs.}
\label{f3}
\end{figure}
More insight on the electronic structure of PdSe$_2$ is achieved by inspecting the ARPES iso-energetic maps. Fig.\ref{f3}a shows the photoemission spectral weight measured on the reciprocal plane at constant energy $E-E_F=-2.3$~eV (in Fig.\ref{f3}b it corresponds to the red horizontal line cutting the band structure). Horizontal and vertical axes are expressed in units of $\pi/a$ and $\pi/b$, respectively. The red dashed lines identify the edges of the orthorhombic BZ. The spectral weight clearly resembles the symmetry of the BZ with peaks (yellow colour) at $\bar{X}$ and $\bar{Y}$ points and valleys (blue colour) at $\bar{\Gamma}$. 
For clarity, each $\bar{\Gamma}$ point has been labeled with the corresponding in-plane Miller indices $(hk)$. Fig.\ref{f3}c shows a photoemission map at $E-E_F = -0.4$~eV, i.e. closer to $E_F$ (in Fig.\ref{f3}b, it corresponds to the blue horizontal line cutting the band structure). The observed elliptical shapes are the surface-projected paraboloids of the top-most VB. By comparing Fig.\ref{f3}a and Fig.\ref{f3}c, one can deduce that all $\bar{\Gamma}$ points with Miller indices satisfying the relation $h+k = $ {\it odd} are missing in panel c. Although surprising at first glance, this observation finds a straightforward rationale by assuming a different crystal symmetry. 

We will unfold this concept referring to Fig.\ref{f3}d-f. Panel d recalls the orthorhombic unit cell of PdSe$_2$ and the corresponding BZ. It is known that ARPES measurements of solids probe the so-called surface-projected BZ (SBZ) based on energy and momentum conservation \cite{arpes}. In panel d the SBZ is represented by the orange-shaded area. A closer inspection of the unit cell reveals that Pd atoms arrange on a {\it face-centered} orthorhombic (fco) lattice, as evident in panel f: if chalcogen atoms were absent, the corresponding first BZ would be the one sketched on the right-hand side \cite{bz} (notice the similarity with the BZ of the standard face-centered cubic lattice) \cite{am}. Panel e compares the SBZs of the orthorhombic (panel d) and the fco (panel f) cells with identical lattice parameters. It can easily be verified that the following relations hold among wavevectors: $\bar{\Gamma} \bar{X} = \pi/a$, $\bar{\Gamma} \bar{Y} = \pi/b$, $\bar{\Gamma} \bar{C} = 2\pi/a$, $\bar{\Gamma} \bar{D} = 2\pi/b$.   

Returning to the photoemission map of Fig.\ref{f3}c, the blue dashed lines identify the SBZ of the fco unit cell and each elliptical shape is centered at the $\bar{\Gamma}$ point of the face-centered lattice. Since Pd atoms arrange on an fco lattice, our ARPES analysis suggests that the top-most VB originates predominantly from Pd orbitals with little contribution from Se. 

\begin{figure}[htb]
\includegraphics[width=130mm]{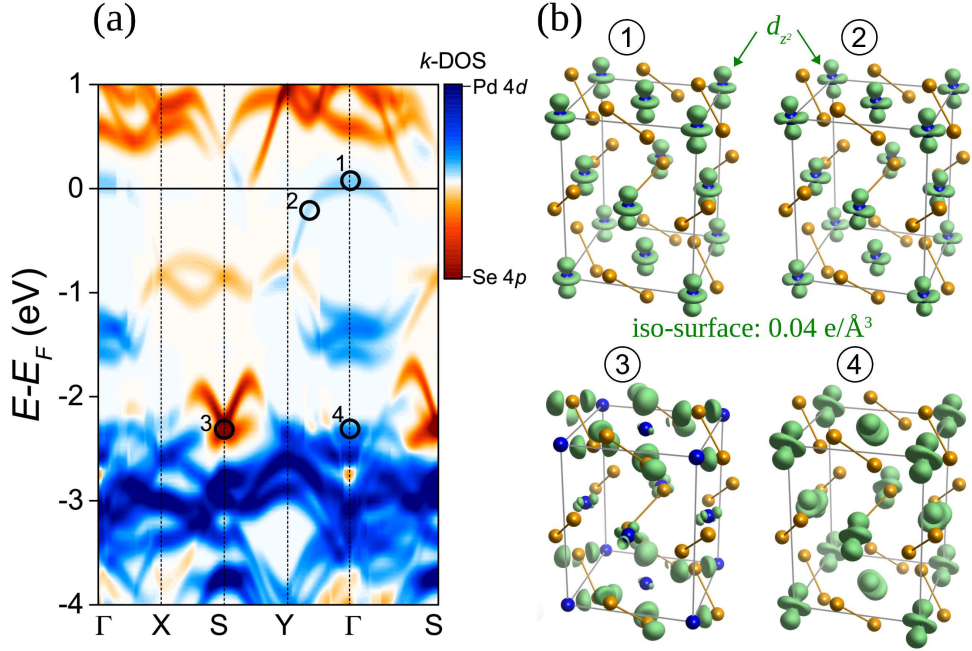}
\caption{(a) Difference between Pd $4d$-projected and Se $4p$-projected $k$-DOS. Blue and red colour represent Pd and Se character, respectively. (b) Electron density (iso-surface value at 0.04~e$/\AA^3$) at the four selected points of the band structure labeled in panel a.}
\label{f5}
\end{figure}
In order to support this hypothesis, we have computed the orbital-projected $k$-DOS and wave functions at selected points of the band structure. Fig.\ref{f5}a reports the {\it difference} between Pd~$4d$-projected and Se~$4p$-projected $k$-DOS along the same path of the reciprocal space employed previously in Fig.\ref{f2} (see Methods for more details on Fig.\ref{f5}). Blue or red colour represents the dominant Pd or Se character, respectively. The highest VB shows a strong Pd~$4d$ character with spectral weight $\sim 3$ times larger than Se~$4p$ and marginal contributions from other atomic orbitals. On the other hand, at $E-E_F = -2.3$~eV (i.e. the same as Fig.\ref{f3}a), the dominant orbital character varies with the specific position on the BZ. Here, it is of interest to visualize the wave functions in real space: Fig.\ref{f5}b displays the electron density at four representative points of the band structure, labeled by circles in panel a. The iso-surface value is taken at 0.04~e$/\AA^3$. On the highest VB (labels 1 and 2) the electron density is centered at Pd atoms with $d_{z^2}$-like symmetry. At $E-E_F = -2.3$~eV (labels 3 and 4) the charge distribution depends on the position within the BZ: at the $S$ point (label 3) it is dominated by Se~$4p$ orbitals \cite{note}, while at $\Gamma$ (label 4) it is combination of Pd~$4d$ states. These results support our hypothesis that the same single orbital of Pd (i.e. $4d_{z^2}$) shapes the top-most VB. Recalling that Pd atoms form an fco lattice, this symmetry is retained also in reciprocal space, as revealed by our ARPES data (see Fig.\ref{f3}c). A simple tight binding approach leads to the same conclusion and is reported in the Methods section. At larger binding energies both Pd and Se orbitals contribute to the band structure, exhibiting the standard orthorhombic symmetry of Fig.\ref{f3}a.      

In conclusion, we have measured the electronic band structure of bulk PdSe$_2$ by angle-resolved photoemission spectroscopy. Within the experimental accuracy, our data confirm its semiconducting nature with a minimum band gap of 50~meV (i.e. the instrumental resolution) since no evidence of conduction band across the Fermi level has been observed, while all electronic dispersive features below $E_F$ are well-reproduced by our DFT calculations. Furthermore, we have demonstrated a remarkable sensitivity of the ARPES technique to site-specific symmetries of the electronic structure. This finding can be pivotal in tuning the electronic properties of PdSe$_2$-based heterostructures \cite{nano}, analogous to the observed dependence of the gap on the band character of MoS$_2$/graphene \cite{nano2,nano3}. Moreover, we envisage that the chemical selectivity of ARPES allows a fine-tuning of the electronic properties. For example, chemical substitution of metal atoms \cite{chem} will give rise to specific changes in the VB related both to doping and to modifications in the surface symmetry, to which ARPES will be sensitive. We believe this implementation is not limited to PdSe$_2$, but it applies to a much wider class of compounds with complex crystal structures and can help clarify the subtle interactions related to correlated electronic phases, such as metal-insulator transitions, charge density waves and superconductivity \cite{cdw1,cdw2}.

{\bf Acknowledgments}
\\
EC acknowledges funding from the Italian PRIN Grant 2017BZPKSZ. Some of the computational work was performed on the University of Bath’s High Performance Computing Facility.

{\bf Author Contributions}
\\
E.C. and C.J.S. proposed the experiments. M.C. performed the ARPES measurements. M.C., C.J.S and E.C. analysed the experimental data. D.W. and E.C. performed the DFT calculations. E.C. wrote the manuscript with input from all authors.

{\bf Competing interests}
\\
The authors declare no competing interests.
\\
\\
{\bf METHODS}

{\bf ARPES.} 
Bulk PdSe$_2$ crystals were obtained commercially from HQ Graphene \cite{hq}. Samples were cleaved in ultra-high vacuum ($10^{-10}$~mbar) to expose a clean surface. Photoemission measurements were performed at the Spectromicroscopy beamline of the Elettra light source using 74~eV linearly polarized radiation focused to an approximately 0.6~$\mu$m diameter spot by a Schwarzschild objective \cite{elettra} and incident at 45$^\circ$ with respect to the sample. The energy and momentum resolution of the hemispherical electron analyzer are $\sim 50$~meV and 0.03~\AA$^{-1}$, respectively. The sample temperature was maintained at 95~K. 

{\bf DFT.} 
The electronic band structure of PdSe$_2$ was computed with the Quantum Espresso package \cite{qe}. Exchange-correlation was considered using the Perdew-Burke-Ernzerhof functional revised for solids (PBEsol). Van der Waals interaction among PdSe$_2$ layers was included using the semiempirical Grimme's DFT-D2 correction \cite{grimme}. Atoms were allowed to relax until the residual forces were below 0.0026~eV/\AA. Cutoff energy of 60~Ry and $8\times 8 \times 6$ $k$-point mesh were used. The iso-surface rendering in Fig.\ref{f5}b was performed with the VESTA software \cite{vesta}.

{\bf Absence of CB evidence in ARPES data.} 
Fig.\ref{s3} reports three constant-energy ARPES maps of PdSe$_2$ across the Fermi level ($E-E_F =+0.5$, 0, $-0.5$~eV, top to bottom). Each map employs a logarithmic intensity scale ranging from 5\% to 100\% of the respective maximum. 
Even well above $E_F$ we have detected no sign of electron pockets originating from the CB, as DFT calculation would predict. Although we are not able to determine the value of the band gap with static ARPES, our data uphold the semiconducting nature of bulk PdSe$_2$.

{\bf Orbital character of the band structure (Fig.\ref{f5}a).} 
\begin{figure}[htb]
\includegraphics[width=130mm]{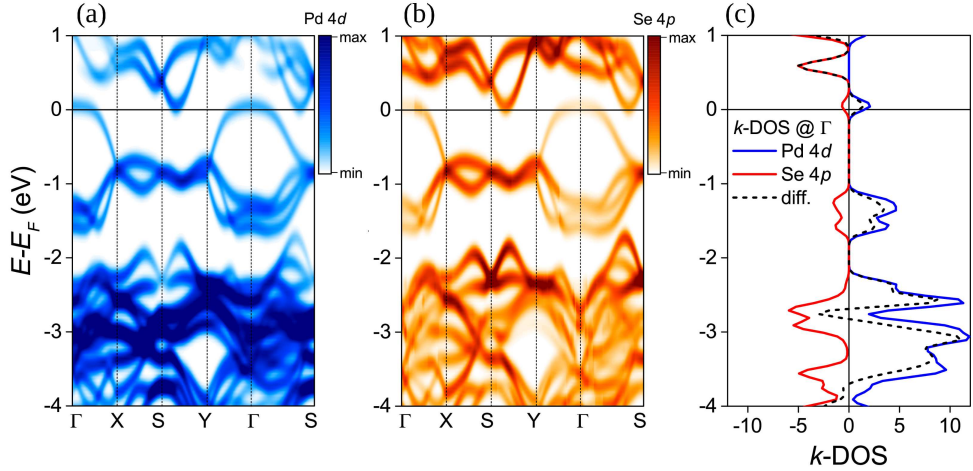}
\caption{$k$-DOS of (a) Pd~$4d$ orbitals and (b) Se~$4p$ orbitals. (c) Comparison of Pd~$4d$ (blue) and Se~$4p$ (red) $k$-DOS at the $\Gamma$ point and their difference (black-dashed).}
\label{s2}
\end{figure}
Pd~$4d$ and Se~$4p$ electrons determine the valence and conduction states of PdSe$_2$. It is instructive to "visualize" the specific orbital character and in particular the Pd-Se {\it duality} along high symmetry lines of the BZ. 
Here, we employ a simple colour-coded approach: the $k$-resolved density of states projected on Pd~$4d$ and Se~$4p$ are shown in Fig.\ref{s2}a and Fig.\ref{s2}b, respectively. Taking the difference between the data of these two graphs we obtain Fig.\ref{f5}a, where blue and red colours encode positive (Pd) and negative (Se) values. As an example, the Pd and Se $k$-DOS (the latter is represented on the negative abscissa) and their  difference at the $\Gamma$ point are shown in Fig.\ref{s2}c. Notice, in particular, that the Pd-projected $k$-DOS at the Fermi level (i.e. the VB top) is approximately 3 times larger than Se, as claimed in the main manuscript.     

\begin{figure}[htb]
\includegraphics[width=130mm]{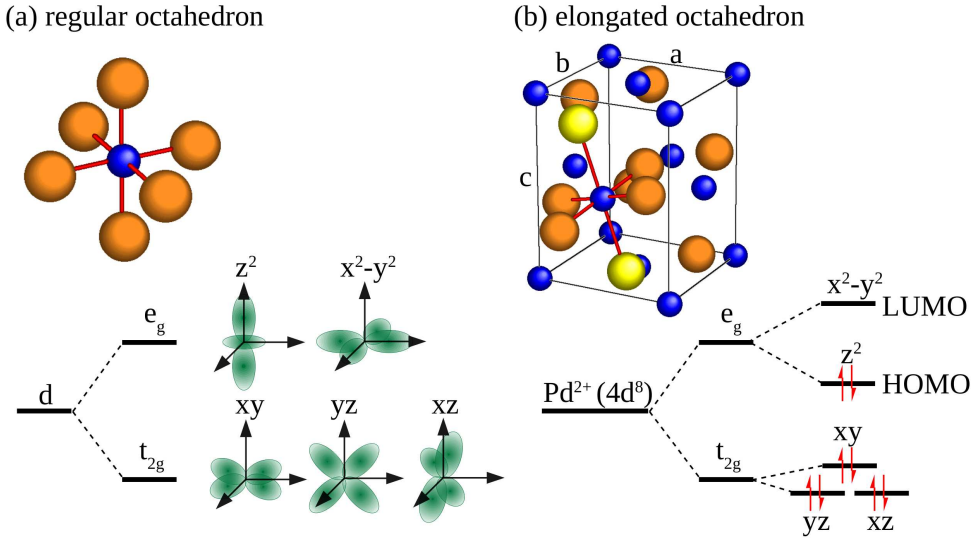}
\caption{(a) Regular MX$_6$ octahedron with the corresponding crystal field splitting of $d$ orbitals into $e_g$ and $t_{2g}$ states. (b) Elongated PdSe$_6$ octahedron along the c-axis: the degenerate $e_g$ states split into HOMO and LUMO.}
\label{s0}
\end{figure}

{\bf Tight binding approach and matrix element effect.}
In a regular MX$_6$ octahedral complex ($O_h$ symmetry) the five outer $d$ orbitals of the transition metal M arrange into the high-energy, double-degenerate $e_g$ and the low-energy, triple-degenerate $t_{2g}$ states \cite{cf}, as sketched in Fig.\ref{s0}a. 
A closer look at the crystal structure of PdSe$_2$, shown in Fig.\ref{s0}b, reveals that each Pd is surrounded by four Se atoms belonging to the same monolayer and two apical atoms (in yellow colour) belonging to the nearest upper and lower layers. The six chalcogen atoms form an octahedron elongated along the $c$-axis ($D_{4h}$ symmetry). This distortion lifts the degeneracy of the $e_g$ states, resulting in the $d_{z^2}$ orbital being energetically more favorable than the $d_{x^2-y^2}$ \cite{cf,cf2,ec1}. As we already pointed out, in PdSe$_2$ the oxidation state of Pd is $+2$ and its electronic configuration is therefore $4d^8$: six of these electrons fill the $t_{2g}$ states completely, while the remaining two occupy the $d_{z^2}$ orbital, leaving $d_{x^2-y^2}$ empty. $d_{z^2}$ is therefore the highest occupied molecular orbital (HOMO) forging the top of the VB, while $d_{x^2-y^2}$ represents the lowest unoccupied molecular orbital (LOMO) contributing the bottom of the CB, in agreement with our own calculations (Fig.\ref{f5}b) and other recent work \cite{ec2}. 

\begin{figure}[htb]
\includegraphics[width=130mm]{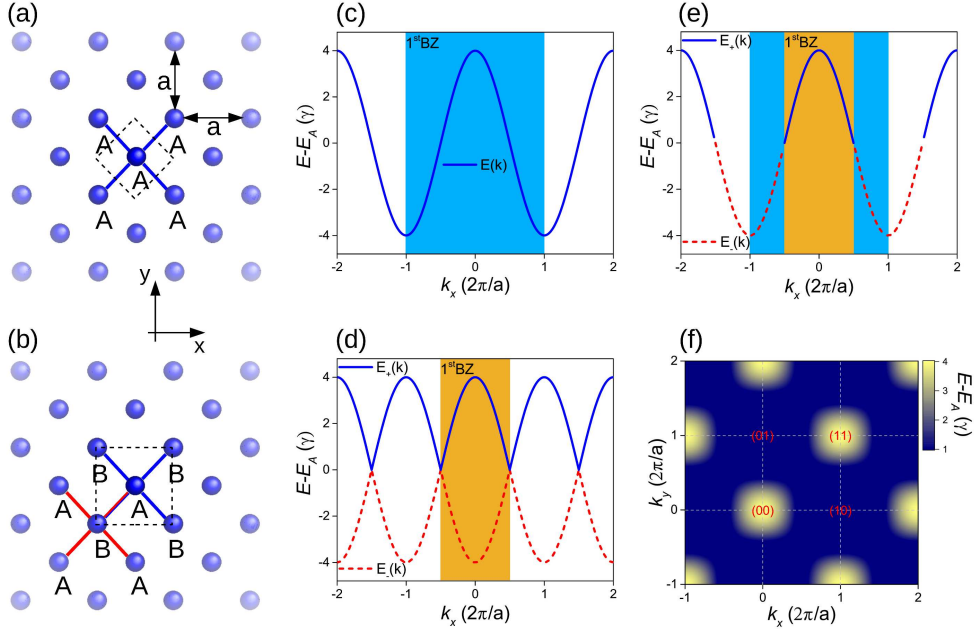}
\caption{Pd atoms of a single PdSe$_2$ layer using (a) the fco unit cell, (b) the orthorhombic cell. Tight-binding VB dispersion employing (c) the fco cell, (d) the orthorhombic cell. (e) Equivalence of (c) and (d) after considering matrix element effect. (f) Top-VB simulation on the $xy$ reciprocal plane (compare with the experimental data of Fig.\ref{f3}c). The Miller indices refer to the orthorhombic cell: the missing VB peaks at (10) and (01) endorse the fco symmetry.}
\label{s1}
\end{figure}
We can now elucidate the symmetry features of the top-VB observed by photoemission using a simple 2D tight binding approach that involves only the HOMO. Metal atoms of a PdSe$_2$ monolayer arrange on a rectangular lattice as shown in Fig.\ref{s1}.
For simplicity we will assume a square lattice of side $a$. If the fco unit cell is used (panel a), a single site $A$ is sufficient to describe the crystal. On the other hand, if the orthorhombic unic cell is assumed (panel b) two sites $A$ and $B$ must be considered: each atom $A(B)$ is surrounded by four nearest neighbors $B(A)$. In the fco case, a tight binding approach is straightforward. Let $|R_A \rangle$ be the Wannier wave function at the specific lattice site $R$ (i.e. the $d_{z^2}$ orbital). The ovelap integral is $\gamma = \langle R_A|U|0_A\rangle$ ($U$ is the periodic lattice potential), where $R$ runs over the 4 nearest neighbors $(+a/2, +a/2)$, $(+a/2, -a/2)$, $(-a/2, +a/2)$, $(-a/2, -a/2)$ and the eigenvalue of the hamiltionian $\mathbb{H}$ is $E(k) = E_A+\gamma\sum_{n.n.}e^{i k R} = E_A+4\gamma\cos(k_x a/2)\cos(k_y a/2)$, with $E_A = \langle R_A|\mathbb{H}|R_A\rangle$. Fig.\ref{s1}c shows the resulting VB dispersion along the $x$-axis ($k_y = 0$). If the orthorhombic cell is used, two Wannier wave functions $|R_A \rangle$ and $|R_B \rangle$ form the basis of the tight binding hamiltonian: 
\begin{eqnarray}\label{tens}
\mathbb{H} &=& \begin{pmatrix} \ E_A & h \\ h^* & E_B  \end{pmatrix} 
\end{eqnarray} 
It can easily be verified that the hopping term between sites $A$ and $B$ is $h = 4\gamma\cos(k_x a/2)\cos(k_y a/2)$ like in the previous fco case. Since sites A and B are equivalent, it also follows that $E_A = E_B$. Thus, the two eigenvalues are $E_\pm(k) = E_A \pm |4\gamma\cos(k_x a/2)\cos(k_y a/2)|$. Fig.\ref{s1}d depicts $E_\pm(k)$ along the $x$-axis ($k_y = 0$). The corresponding eigenstates are $(\pm\frac{h}{|h|},1)$ and the generic wave function at the lattice site $R$ reads \cite{matrix}: $|R_\pm \rangle = \pm\frac{h}{|h|} e^{i k R_A}|R+R_A\rangle + e^{i k R_B}|R+R_B\rangle$. In the free electron final state approximation (here, the final state $|k_f \rangle$ is a plane wave with wavevector $k_f$), the photoemission matrix element $\mathbb{M}$ is expressed as the Fourier component of the tight binding orbital $|0_\pm \rangle$ \cite{matrix}, i.e.
\begin{eqnarray}
\mathbb{M} \propto \langle k_f|0_\pm \rangle &=& \langle k_f|(\pm\frac{h}{|h|} e^{i k R_A}|R_A\rangle + e^{i k R_B}|R_B\rangle)\\
&=& \pm\frac{h}{|h|} e^{i k R_A}\langle k_f|R_A\rangle + e^{i k R_B}\langle k_f|R_B\rangle\\
&=&\pm\frac{h}{|h|} e^{i (k-k_f) R_A}\langle k_f|0_A\rangle + e^{i (k-k_f) R_B}\langle k_f|0_B\rangle
\end{eqnarray}
which, using the momentum conservation ($k=k_f$), leads to the following photoemission intensity: 
\begin{widetext}
\begin{equation}
I_\pm \propto \left| \mathbb{M} \right|^2 \propto
\left|\pm\frac{h}{|h|}\langle k_f|0_A\rangle + \langle k_f|0_B\rangle \right|^2 =  \left|\langle k_f|0_A\rangle \right|^2+\left|\langle k_f|0_B\rangle \right|^2 \pm\frac{h}{|h|} 2 \Re (\langle k_f|0_A\rangle \langle 0_B|k_f\rangle)
\end{equation}
\end{widetext}
that simplifies to
$I_\pm \propto 2\left|\langle k_f|0_A\rangle \right|^2\left(1 \pm\frac{h}{|h|}\right)$ since $|0_A\rangle = |0_B\rangle$. At this point we notice that $\frac{h}{|h|} = $ sgn$[\cos(k_x a/2)\cos(k_y a/2)]$. Thus, $I_\pm \propto (1 \pm $ sgn$[\cos(k_x a/2)\cos(k_y a/2)]$. Referring to Fig.\ref{s1}d, it is easily verified that the previous equation completely suppresses the photoemission intensity of one eigenvalue $E_\pm(k)$ depending on the values of $k_x$ and $k_y$. Fig.\ref{s1}e-f show the results: as expected, the fco band structure of Fig.\ref{s1}c and the experimental data of Fig.\ref{f3}c are recovered since the use of two equivalent sites is redundant and the appropriate unit cell is the fco.


\begin{thebibliography}{1000}
\bibitem{note}
Point 3 in Fig.\ref{f5}a is four-fold degenerate. Here we show the iso-surface of one of the bands, the others differing only by $4p$ orbital orientation and/or Se atom.
\bibitem{cf2}
If ligands with apical atoms are negligible, $d$ orbitals split into the square planar coordination where $d_{xy}$ has higher energy than $d_{z^2}$ (and the former becomes the HOMO), see Ref.\cite{cf}.
\end{thebibliography}
\end{document}